\documentclass[final,english]{bullsrsl}[2022/06/15]
\usepackage[latin1]{inputenc}
\usepackage[T1]{fontenc}

\usepackage{natbib} 
\usepackage{graphicx} 
\usepackage{amsmath,amsthm,amssymb,amsfonts,fancyhdr}
\newcommand{\as}{^{\prime\prime}}

\def\be{\begin{equation}}
\def\ee{\end{equation}}

\usepackage[acronym]{glossaries}
\usepackage{xcolor}
\usepackage[linesnumbered,ruled,vlined]{algorithm2e}
\SetKwComment{Comment}{$\triangleright$\ }{}
\SetCommentSty{itshape}
 
\begin{document}
 
\title{Deep Photometry of Suspected Gravitational Lensing Events: Potential Detection of a Cosmic String}

\author[affil={1}, corresponding]{Margarita}{Safonova}
\author[affil={2}]{Igor~I.}{Bulygin}
\author[affil={2}]{Olga~S.}{Sazhina}
\author[affil={2}]{Mikhail~V.}{Sazhin}
\author[affil={3}]{Priya}{Hasan}
\author[affil={1}]{Firoza}{Sutaria}
\affiliation[1]{Indian Institute of Astrophysics (IIA), Bangalore, India}
\affiliation[2]{Sternberg Astronomical Institute (SAI), Moscow, Russia}
\affiliation[3]{Maulana Azad National Urdu University, Hyderabad, India}
\correspondance{margarita.safonova@iiap.res.in}
\date{}

\maketitle

\begin{abstract}
Cosmic strings (CS) are one-dimensional cosmological-size objects predicted in realistic models of the early Universe. Analysis of the cosmic microwave background (CMB) anisotropy data from the  Wilkinson Microwave Anisotropy Probe (WMAP) and Planck surveys revealed several CS candidates. One of the candidates, CSc-1, was found to be most reliable because of the statistically significant chains of gravitational lensing (GL) candidates in its field. We  observed the brightest of the objects in the CSc-1 field, a galaxy pair SDSSJ110429.61+233150.3. The significant correlation between the spectra of the two components indicates the possible GL nature of the pair. Our simulations of observational data in the CSc-1 field shows that a large number of pairs can be explained by the complex geometry of the CS. Simulations of the SDSSJ110429 galaxy pair has shown that the observed angle between the components of the pair can be explained if the CS is strongly inclined and, possibly, bent in the image plane. In our preliminary data, we also detected the sign of the sharp isophotal edge in one image, which along with CMB and spectral data strongly suggests the possibility of a CS detection. 
\end{abstract}

 
\section{Introduction}

Cosmic strings (CS) are one-dimensional extended objects predicted by modern cosmology. Till now there have been no confirmed detections of these objects. CS make a global conical geometry of the Universe. CS can be either topological (Abelian-Higgs, etc.) or formed as a result of the interaction of multi-dimensional spaces (F- and D-strings). In the latter case, CS could serve as a unique proof of higher-dimensional theories. 

There are two primary methods of observational searches of CS. The first one is based on the analysis of the CMB anisotropy (WMAP, Planck data) with an appropriate set of step-like filters. The second one is based on the search of special chains of GL events which could be produced by CS, such as e.g. galaxy pairs. Using both these methods, the most reliable CS candidate was identified -- CSc-1 \citep{olga}. Using the SDSS survey, about 30 objects in CSc-1 field were found to be possible GL images in optical bands, based on their morphological and photometric characteristics \citep{review}. These objects are in the range of 19.7--23 magnitudes, and thus require large stable telescopes for successful observations (such as e.g. DOT), where high-signal-to-noise (SNR) images can help identify the morphological details characteristic only to CS lensing; including the so-called isophote sharp edges arising due to step-like CS-nature \citep{sazhin2007}. We performed photometric and spectroscopic observations of the brightest object in this chain -- a galaxy pair SDSSJ110429.61+233150.3 (SDSSJ110429-A,B henceforth). The results of the observations and modelling strongly suggest the possibility of a CS detection.

\section{Observations}

The CS lens candidate SDSSJ110429-A,B consists of two galaxies situated at an angle to each other with a distance of about $4.55^{\as}$ between the cores (Fig.~\ref{images}). Photometric and spectroscopic observations were obtained using the Himalayan Faint Object Spectrograph and Camera (HFOSC) mounted on the 2.0-m Himalyan Chandra Telescope (HCT) of the Indian Astronomical Observatory (IAO), located at 4500 m above sea level. HFOSC is equipped with a Thompson CCD of 2048$\times$2048 pixels with a pixel scale of $0.296^{\as}$/pix, equivalent to a total field of view (FOV) of 10$'\times$$10'$. The readout noise, gain and readout time of the CCD are 4.87 $\bar{e}$, 1.22 $\bar{e}$/ADU and 90 sec, respectively. Photometry was performed in Bessel R-band with variable exposures. In addition, we have obtained several open filter frames of the object. Since the object is at a high redshift, spectroscopy was carried out using the Gr8/167l configuration, with a bandpass of 5800-8350 \AA, a resolution of $\lambda/{\Delta \lambda} = 2190$ and a slit width of $1.92\as$. Spectroscopy was done in single exposures with the slit passing through the cores of both galaxies.

The photometric data was subjected to the usual image reduction process (bias subtraction, flat-fielding, and cosmic rays removal) using IRAF \citep{iraf} scripts developed by our group. Flat fields were constructed by taking a median of multiple, dithered images of the twilight sky, thus removing any contamination from cosmic rays and stellar sources. Illumination and fringe corrections were not required due to the small FOV, and the dark current is negligible since the CCD is cooled down to $-100^\circ$C. The spectroscopic exposures were preprocessed, cleaned of cosmic rays, and the spectra were extracted using standard IRAF procedures. Spectra were not flux-calibrated as we only needed the wavelength calibration for the current task.

\section{Analysis}

\subsection{Properties of images}

\begin{figure}[h!]
\centering
\includegraphics[width=0.38\textwidth]{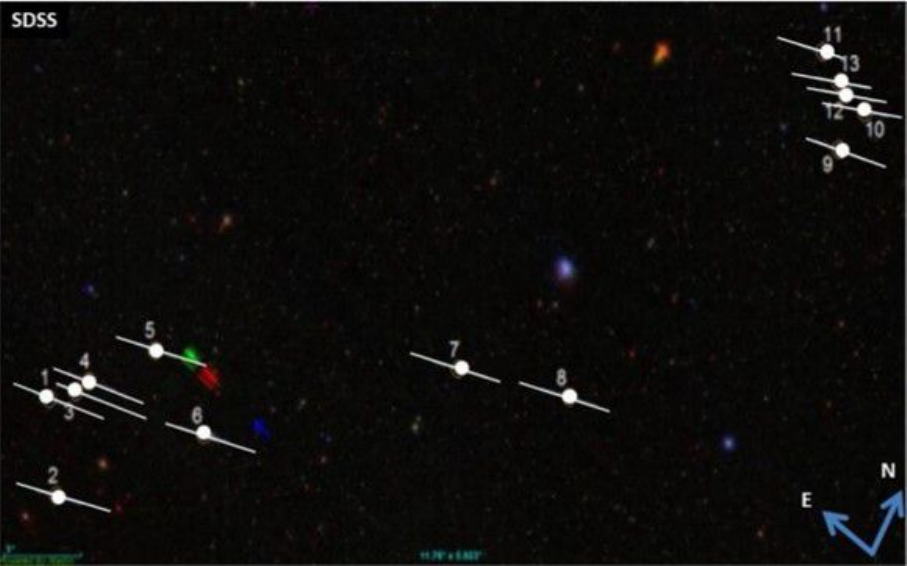} 
\includegraphics[width=0.59\textwidth]{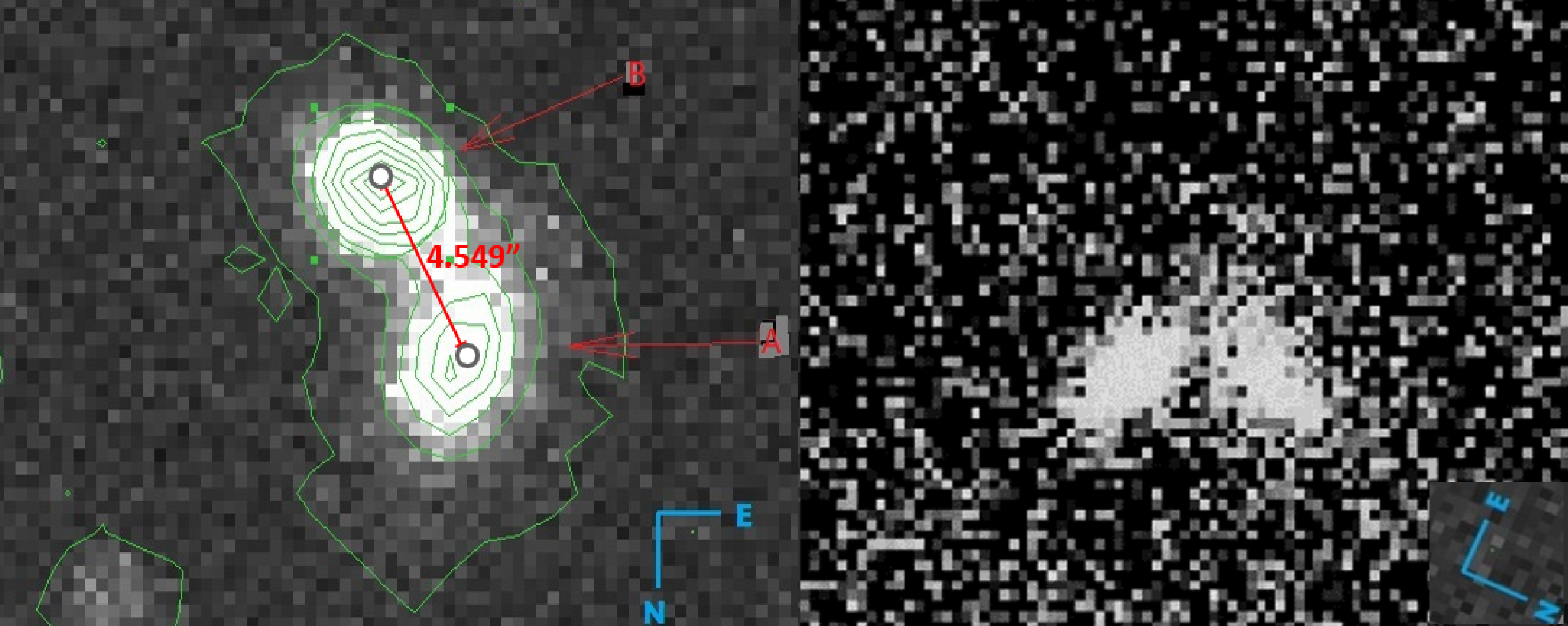}
\begin{minipage}{12cm}
\caption{{\it Left}: \small A portion of the CSc-1 field: white circles indicate galaxy pairs with angular distances 4$^{\as}$--6${\as}$ and white lines show the expected orientation of the CS \citep{olga}. {\it Middle}: HCT 1800-sec $R$-band image of object No.9 -- galaxy pair SDSSJ110429-A,B. {\it Right}: HCT 120-sec open filter exposure showing the indication of of a sharp edge.}
\end{minipage}
\label{images}
\end{figure}

The instrumental $r$ magnitudes were converted to the $R$ standard system using 5 photometric standard stars from a standard field PG0942-029 using photometric transformation equation,
\be
R = -0.723219 + 0.987765 \cdot r_{\rm corr}\,,\nonumber 
\ee
where $r_{\rm corr}$ is instrumental magnitude, normalized to 1800-sec exposure time, $r_{\rm norm} = r + 2.5 \cdot \log(T_{\rm exp})$, 
and corrected for airmass and extinction $r_{\rm corr} = r_{\rm norm} - k_R \cdot X_R$, where $X_R = 1.04735$ is airmass and $k_R = 0.09$ is the $R$-band extinction coefficient at HCT \citep{Stalin}. The resulting standard $R$ magnitudes of the candidates are $19.754 \pm 0.019$~(A) and $19.646 \pm 0.0197$ (B). A frame taken in {\em spec} mode (400$\times$400-pix open filter) indicates the sign of the isophote sharp edge -- a signature of the CS extended source lensing (Fig.~\ref{images}, {\it Right}).

\subsection{Properties of spectra}

\begin{figure}[h!]
    \centering
    \includegraphics[width=0.3\textwidth]{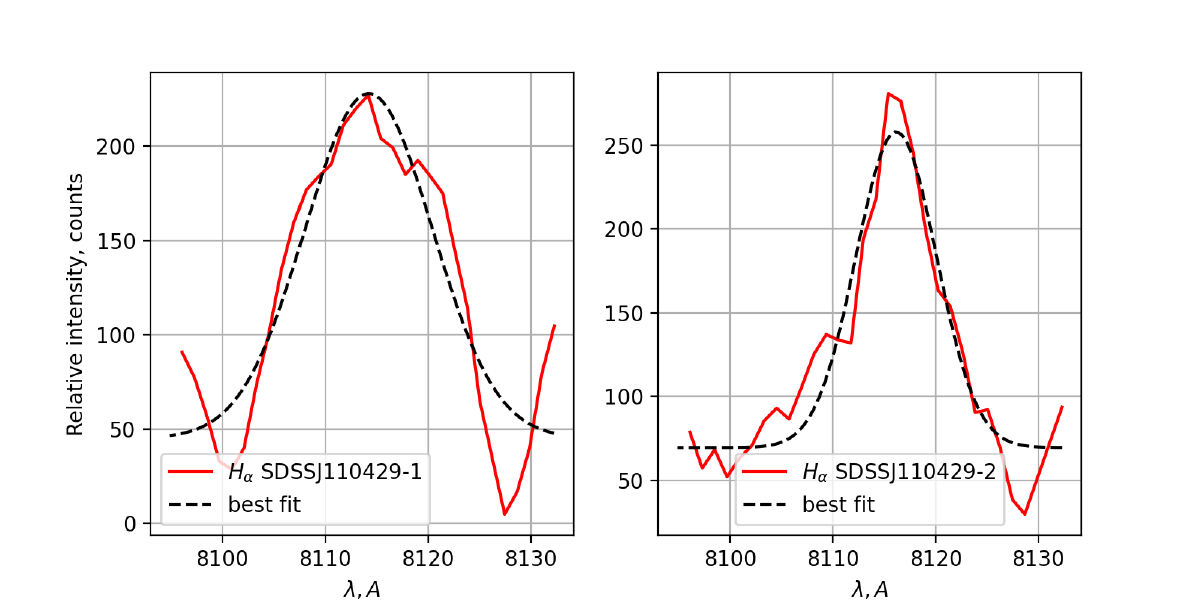}
    \includegraphics[width=0.3\textwidth]{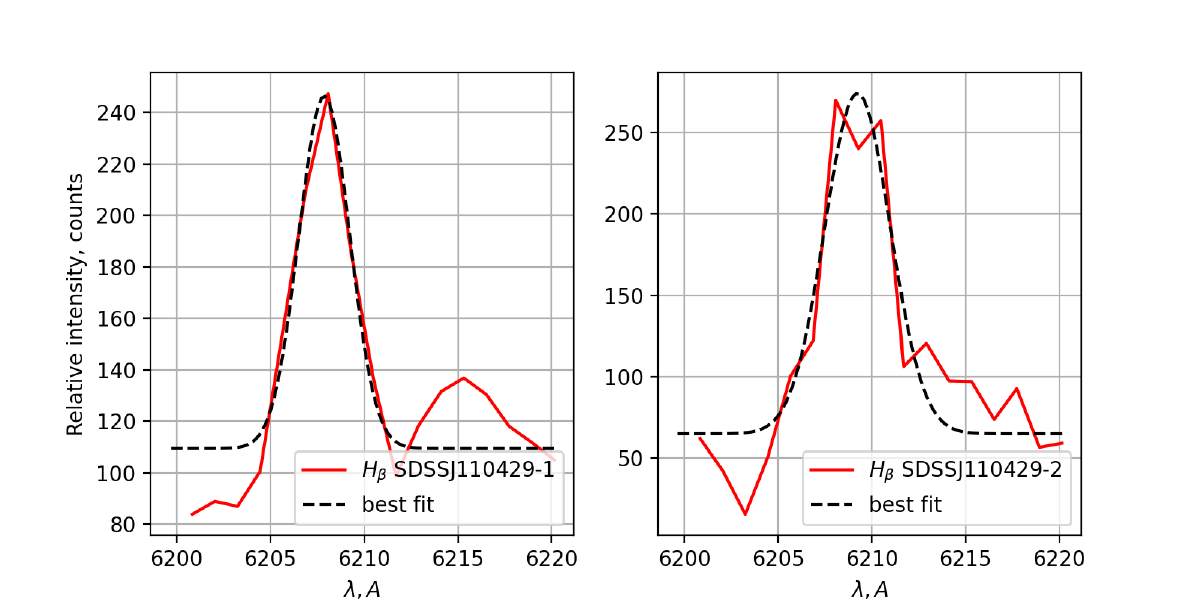}
    \includegraphics[width=0.3\textwidth]{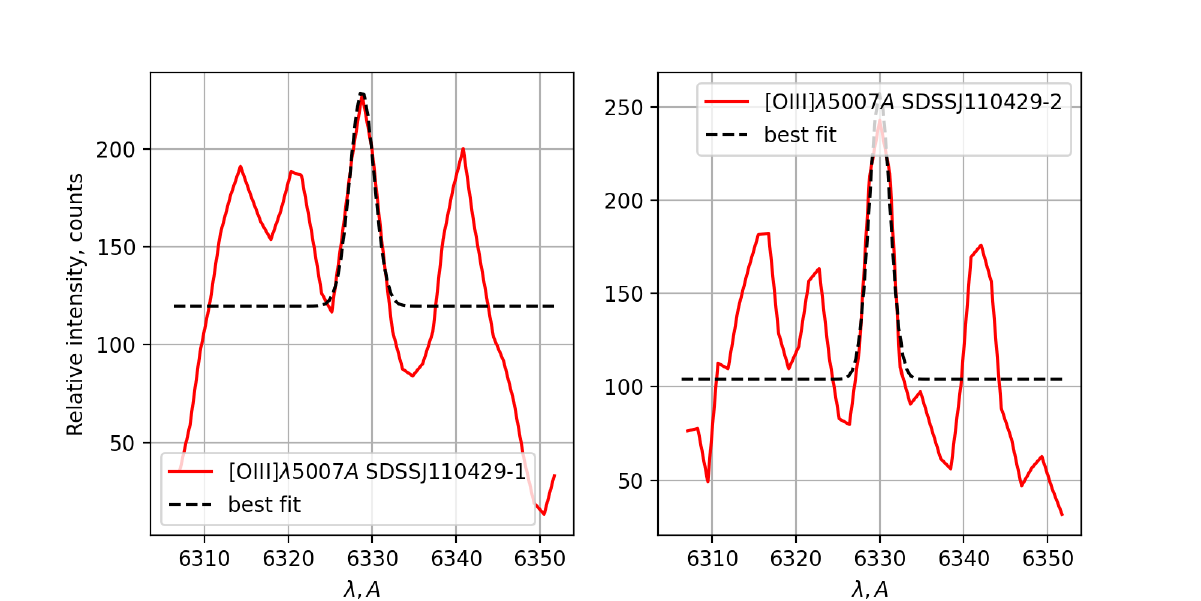}
    \includegraphics[width=0.3\textwidth]{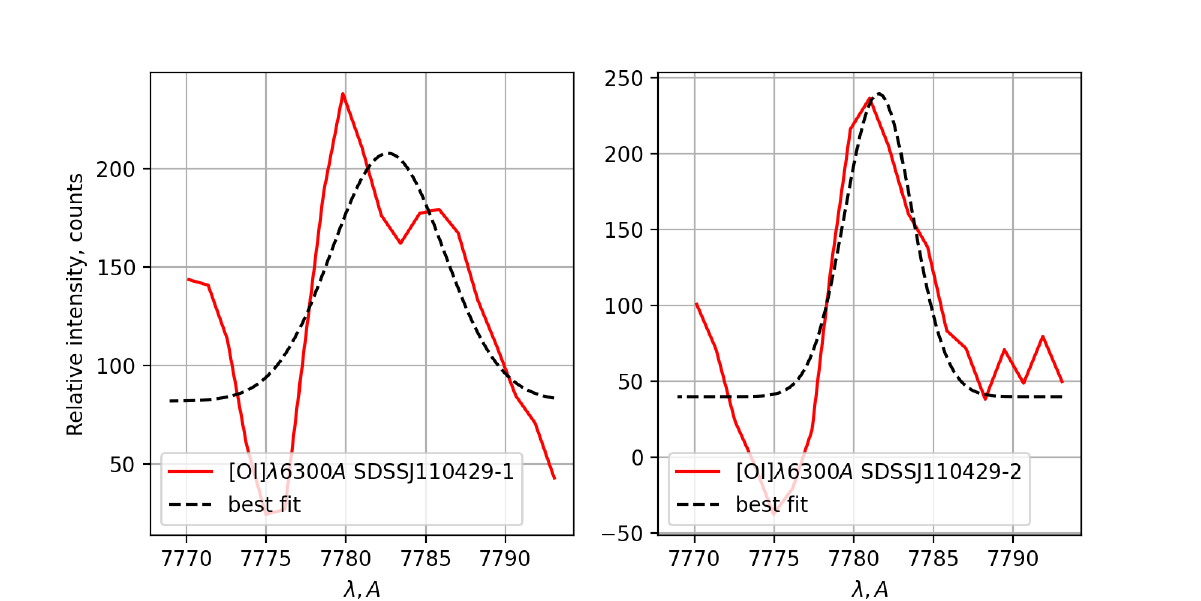}
    \includegraphics[width=0.3\textwidth]{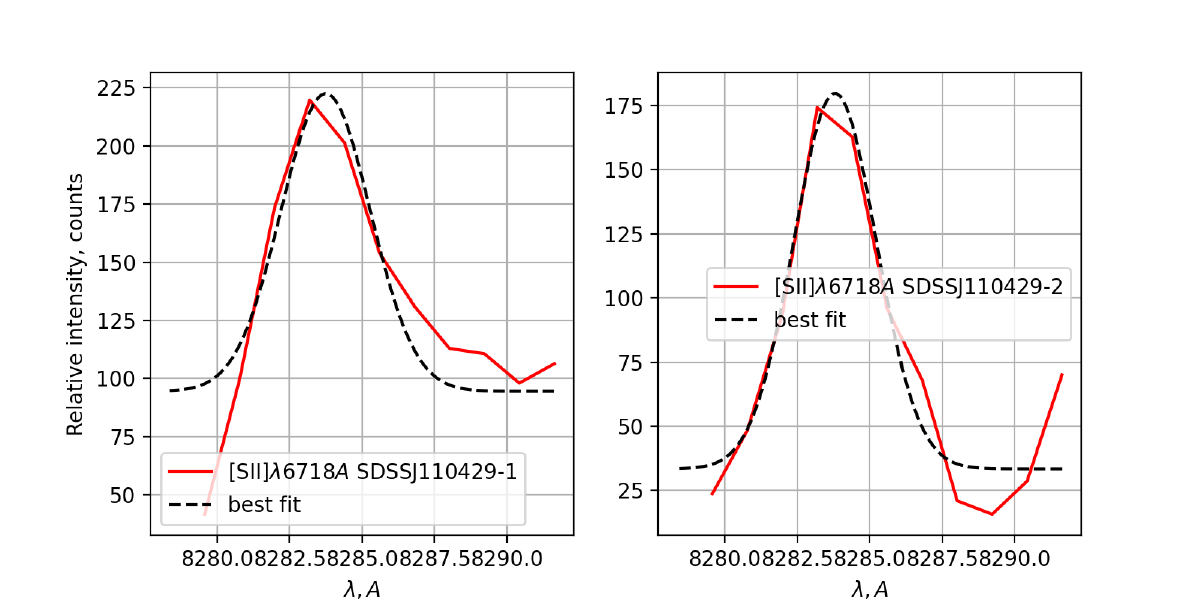}
    \includegraphics[width=0.3\textwidth]{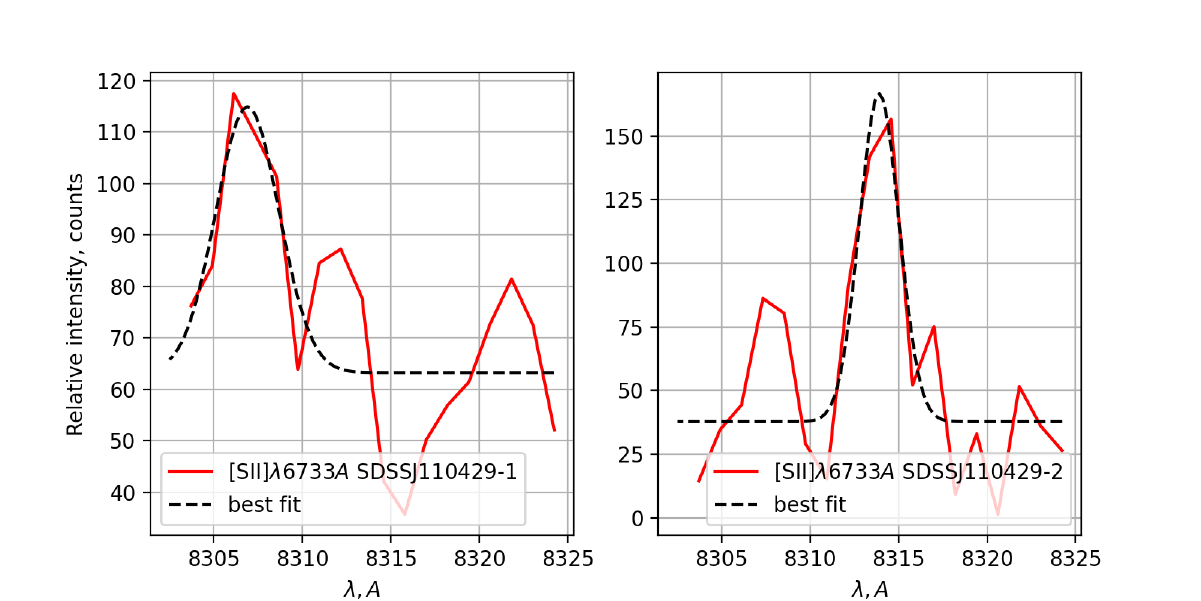}
    \includegraphics[width=0.3\textwidth]{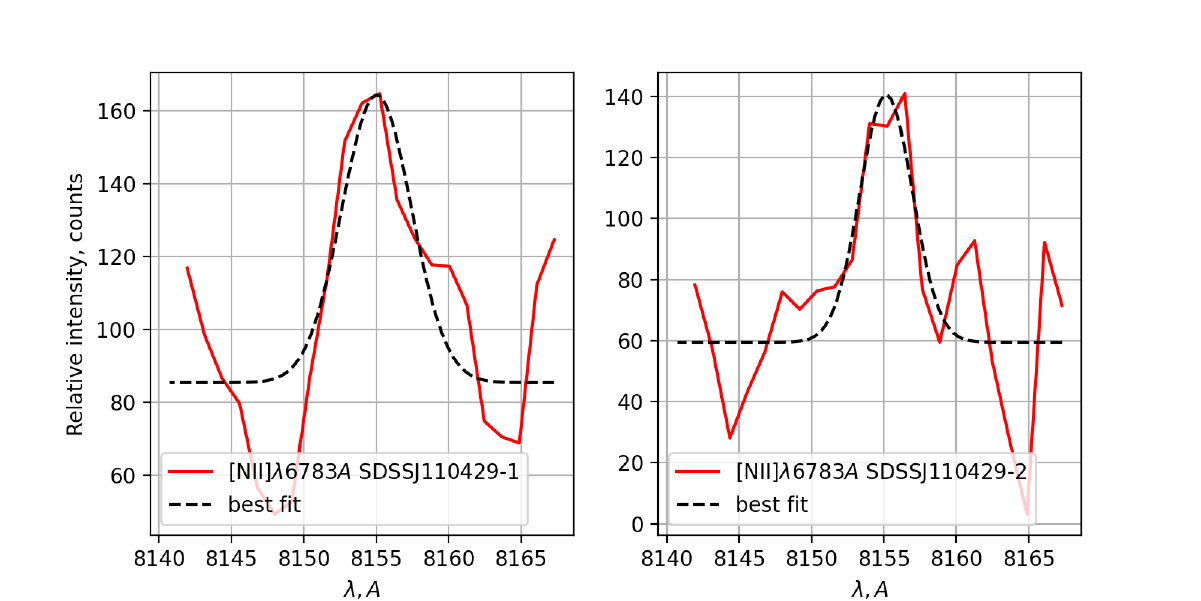}
\begin{minipage}{12cm}
\caption{\small Best fits for the detected lines in spectra of both galaxies. In each pair of plots, the left one corresponds to SDSSJ110429-B, and the right one corresponds to SDSSJ110429-A.}
    \label{lines}
\end{minipage}    
\end{figure}

If these two galaxies are GL images, the spectra of both objects should be identical or, at least, be linearly dependent. Significance of the lensing hypothesis was tested by the following criteria: (a) correlations between the spectra, and (b) $\chi^2$ criterion to compare the properties of individual lines. We calculated Pearson, Spearman and Kendall correlations between the smoothed spectra on the scale of $\delta \lambda = 3.6$ \AA \:, corresponding to the spectral resolution. The correlation coefficients are $0.571$, $0.613$ and $0.447$, respectively. 
\begin{figure}[h!]
\centering
\includegraphics[width=1.\textwidth]{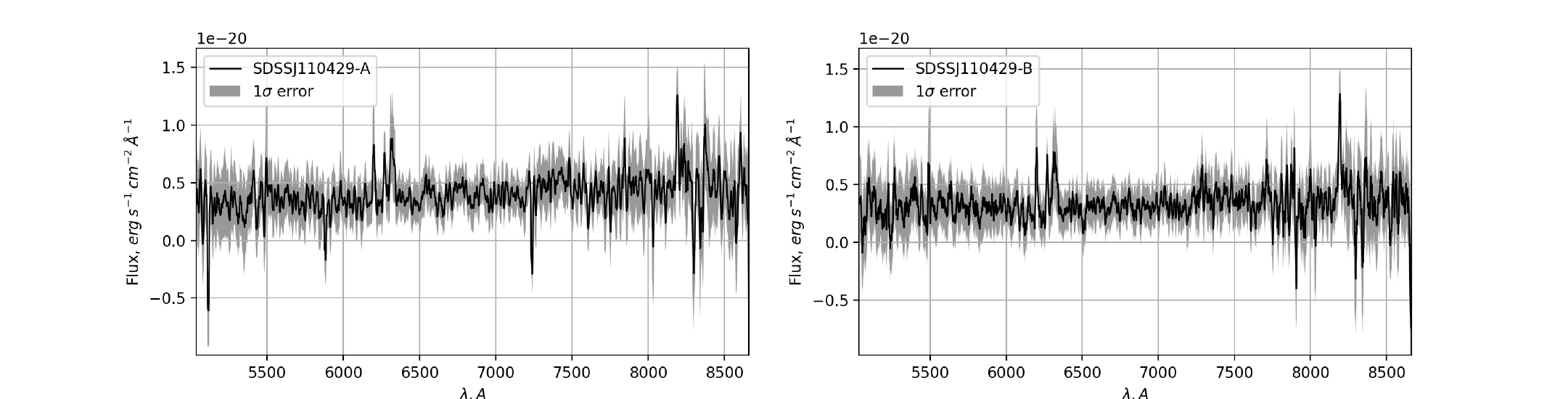} 
\begin{minipage}{12cm}
\caption{\small Spectra of both galaxies smoothed with rolling mean on a scale $\delta \lambda= 3.6$ \AA.}
\label{spectra}
\end{minipage}
\end{figure}
\noindent
Several lines (H$_\alpha$, H$_\beta$, [OIII]$\lambda$5007, [NII]$\lambda$6583, [SII]$\lambda$6718, [SII]$\lambda$6733 and [OI]$\lambda$6300) were identified and fitted by a Gaussian profile (Fig.~\ref{lines}). In order to analyze the properties of spectra without the noisy continuum, $\chi^2$ criteria was calculated for (1) profiles of strongest lines H$_\alpha$, H$_\beta$ and [OIII]$\lambda$5007, and (2) for the widths of all lines. The results are the following: $\chi^2_1/\text{DOF}_1 = 0.69$, $\chi_2^2 = 10.32, \:\text{DOF}_2 = 7, \:\text{p-value} = 0.9$. This suggests that the spectra are identical. Based on the H$_{\alpha}$ line, the redshift of both components is the same $z=0.236$.

\subsection{Modelling of lensing by the curved cosmic string}
    
\begin{figure}[h!]
\centering
\includegraphics[width=0.9\textwidth]{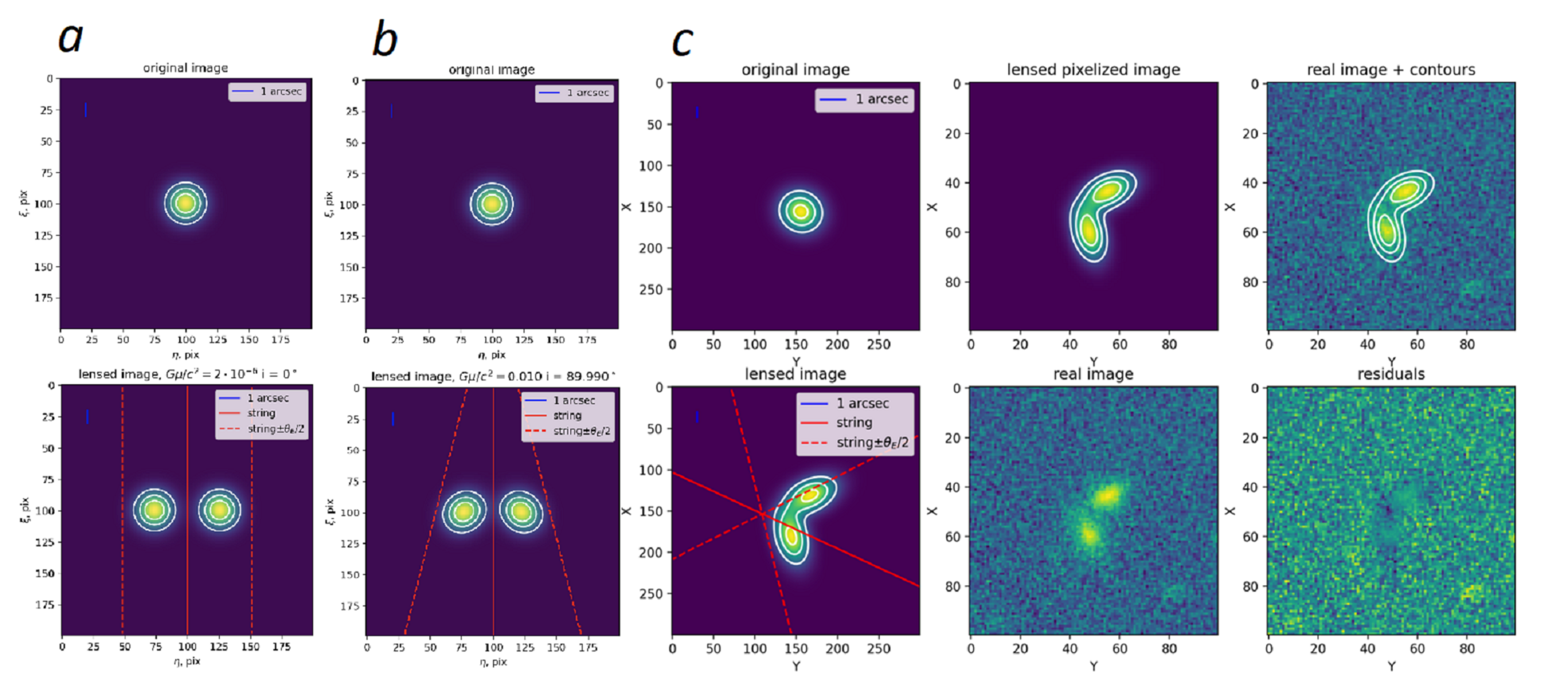}
\begin{minipage}{12cm}
\caption{{\it Left}: \small Modelling of CS lensing on extended source. a) String lying in the image plane. b) String inclined to the image plane. c) The result of LS minimization of inclined lensing model of SDSSJ110429-A,B. $\theta_E$ is essentially the Einstein angle -- maximum angle between the string and the source for lensing to occur. $\theta_E = \Delta\theta (1 - R_s / R_g)$ for a string lying in the image plane, and is a function of $i$ and a coordinate along the string for the inclined one.}
\label{lensed_LS}
\end{minipage}
\end{figure}

According to the theory of lensing on a CS, two identical images are produced (Fig.~\ref{lensed_LS}{\it a}), separated by an angle proportional to the string deficit angle $\Delta \theta = 8\pi G\mu / c^2$, where $\mu$ is the linear density of the CS \citep{sazhin2007}. But various geometric effects can produce a difference in positional angles of images (Fig.~\ref{lensed_LS}{\it b}). We suggest that galaxy pair SDSSJ110429-A,B can be a GL event on a string strongly inclined to the image plane, at inclination angle $i \approx 90^\circ$. Using the least squares (LS) minimization and taking into account the PSF of the image, we have found the string orientation that can produce such an image (see Fig.~\ref{lensed_LS}{\it c}). The LS minimization results are the following: $i = 89.9995^\circ$, $G\mu / c^2 = 0.050$, $R_s / R_g = 0.31$, where $R_{s}$ and $R_{g}$ -- distances from observer to the string and to a lensed galaxy, respectively.
One may argue that the bending of the CS in the image plane can also distort the images. We discuss and consider both effects in the upcoming publication, since it involves calculations using general relativity \citep{theory}.

\section{Conclusions}

Significant correlation between the spectra of the two components indicates the possible GL nature of the pair. Lensed images of background galaxies by galaxies/clusters of galaxies typically produce isophotal distortions due to the non-uniform shear (tidal field). The only known configuration with observed undistorted images is a CS case, because near the CS the space-time is flat. Our modelling of observational data in CSc-1 shows that a large number of pairs can be explained by the complex geometry of the string. Considering a model of a CS with the bend in the image plane can improve the search for GL event candidates. In particular, modeling of the SDSSJ110429-A,B galaxy pair has shown that the observed angle between the components of the pair can be explained if the string is strongly inclined to the line of sight and, possibly, bent in the image plane. In our preliminary data, we also detected the sign of the sharp isophotal edge in one image, which along with CMB and spectral data strongly suggests the possibility of a CS detection. There can be few dozen long strings crossing horizon volume and, therefore, any survey aimed at detecting them needs to be multi-band, very deep, and of high signal to noise (or photometric) sensitivity, achievable by the 4-m class telescopes such as the DOT of ARIES, Nainital.

\begin{acknowledgments}
MS acknowledges the financial support by the DST, Government of India, under the Women Scientist Scheme (PH) project reference number SR/WOS-A/PM-17/2019. IIB acknowledges the financial support by the <<BASIS>> foundation.
\end{acknowledgments}

\begin{furtherinformation}

\begin{authorcontributions}
MS: writing-review and editing. MVS and OSS: Conceptualization (lead), writing-review and editing (equal). MS and FS: observations, data collection and data processing. PH: Software, photometry, formal analysis. IIB: Software, simulations, formal analysis.
\end{authorcontributions}

\begin{conflictsofinterest}
The authors declare no conflict of interest.
\end{conflictsofinterest}
\end{furtherinformation}

\bibliographystyle{bullsrsl-en}
\bibliography{extra}
 
\end{document}